\documentclass[twocolumn,showpacs,floats,floatfix,superscriptaddress,aps,pra]{revtex4-1}

\usepackage{amsfonts}
\usepackage{amssymb}
\usepackage{amsmath}
\usepackage{calc}
\usepackage{graphicx}
\usepackage{bm}

\usepackage[normalem]{ulem}
\usepackage{amsmath,amssymb}
\usepackage{multirow}
\usepackage{xcolor,soul}
\usepackage{srcltx}
\usepackage{hyperref,graphicx}

\def\be{ \begin{equation}}
\def\ee{ \end{equation}}
\def\bea{ \begin{eqnarray}}
\def\eea{ \end{eqnarray}}
\def\bse{ \begin{subequations}}
\def\ese{ \end{subequations}}
\def\bc{ \begin{center}}
\def\ec{ \end{center}}

\begin{document}

\author{Stefano Longhi$^{*}$} 
\affiliation{Dipartimento di Fisica, Politecnico di Milano, Piazza L. da Vinci 32, I-20133 Milano, Italy}
\affiliation{IFISC (UIB-CSIC), Instituto de Fisica Interdisciplinar y Sistemas Complejos, E-07122 Palma de Mallorca, Spain}
\email{stefano.longhi@polimi.it}

\title{Phase transitions and bunching of correlated particles in a non-Hermitian quasi crystal}
  \normalsize


%
\bigskip
\begin{abstract}
\noindent  
Non-interacting particles in non-Hermitian quasi crystals display localization-delocalization and spectral phase transitions in complex energy plane, that can be characterized by point-gap topology.
Here we investigate the spectral and dynamical features of two interacting particles in a non-Hermitian quasi crystal, described by an effective Hubbard model in an incommensurate sinusoidal potential with a complex phase,  and 
unravel some intriguing effects without any Hermitian counterpart.  Owing to the effective decrease of correlated hopping introduced by particle interaction, doublon states, i.e. bound particle states, display a much lower threshold for spectral and localization-delocalization transitions than single-particle states, leading to the emergence of mobility edges. Remarkably, since doublons display longer lifetimes, two particles initially placed in distant sites tend to bunch and stick together, forming a doublon state in the long time limit of evolution, a phenomenon that can be dubbed {\em non-Hermitian particle bunching}.
 \end{abstract}

\maketitle

\section{Introduction}
Topological phases, localization and novel phase transitions in non-Hermitian systems with periodic or aperiodic order have recently sparked a great interest in a wide variety of physical systems, ranging from condensed matter physics to cold atoms and classical systems (see e.g. \cite{r1,r2,r3,r4,r5,r5b,r5c,r5d,r5e,r5f,r5g,r5h,r5i} and references therein). 
 Non-interacting particles in crystalline systems described by an effective non-Hermitian Hamiltonian display a variety of exotic physical effects, such as a non-trivial point-gap topology, the non-Hermitian skin effect, the breakdown of the bulk-boundary correspondence based on Bloch band topological invariants, and a variety of dynamical and transport signatures
 \cite{r6,r7,r8,r9,r10,r11,r12,r13,r14,r15,r16,r17,r18,r19,r20,r21,r22,r23,r24,r25,r26,r27,r28,r29,r30,r31,r32,r33,r33a,r33b,r33c,r34,r35,r36,r36b,r37,r38,r39,r40,r41,r42,r43,r44,r45,r45b,r46,E1,E2,E3,E4,E5,E6,r47,r47a,r47b,r47c,r47d,r47e,
 r47f,r47g,r47h,r47hh,r47i,r47l,r47m,r47n,r47o,r47p,r47pp,r47q,r47r,r47s}.  Recent experimental realizations of synthetic matter with controllable non-Hermitian Hamiltonians using different platforms, such as photonic systems, cold atoms in optical lattices, mechanical and topolectrical systems, have  lead to the observation of such exotic physics.\\ 
 The study of non-Hermitian physics in such systems is being extended into several directions, unraveling a plethora of intriguing effects which do not have any counterpart in Hermitian systems. 
 For example, in systems with aperiodic order (quasi crystals), non-Hermiticity induces phase transitions that are beyond the paradigm of Hermitian quasi crystals \cite{r48,r49,r50,r51,r52,r53,r54,r55,r56,r57,r58,r59,r60,r60a,r60b,r60c,r60d,r60e,r60f,r60g,r60h,r60i,r60l,r60m,r60n}.
Non-Hermitian extensions of the famous Aubry-Andr\'{e} model \cite{A2} have attracted great interest recently, unraveling the rich interplay between disorder, non-Hermiticity and topology. In such systems, the localization-delocalization phase transition and mobility edges in complex energy plane, separating extended and  localized states, can be rather generally characterized by point-gap topological numbers \cite{r1,r2,r3,r4,r5,r5b,r5c,r5d,r5e,r5f,r5g,r5h,r50,r52,r53,r54,r55,r57,r58,r59,r60}.
Another interesting ramification is provided by the many-body physics of non-interacting non-Hermitian systems \cite{MB1,MB2,MB3,MB4,MB5,MB6,MB7,MB8,MB9,MB10,MB11}, where the construction of many-body states involves ramified symmetry classes leading to unique topological phases. Owing to the different lifetimes of single particle eigenstates, non-Hermitian many-particle systems may not
attain an equilibrium state but rather a non-equilibrium steady state at long times.\\ 
  Recently, there is a growing focus on exploring non-Hermitian phenomena in {\em  interacting} many-particle systems, aimed at understanding the interplay between non-Hermitian skin effect, interaction, and non-Hermitian topological phases in correlated systems \cite{I1,I2,I3,I4,I6,I7,I8,I10,I11,I12,I13,I15,I16,I17,I18,I19,I20}. Quantum many-body phases lead to novel manifestations beyond single particle physics,
where the collective behavior of a large number of constituents offers several exotic phases of
matter. Among the simplest and intriguing phenomena in strongly
correlated quantum systems of both fermions and bosons is
the formation of doublons, i.e. pairs of bound
particles occupying the same lattice site \cite{DB1,DB2,DB3,DB4,DB5,DB6,DB7,DB8}.  
Such states are readily found within the standard Fermi-Hubbard or Bose-Hubbard models in the two-particle sector of Hilbert space. 
For sufficiently strong interactions, either attractive or repulsive,
isolated doublons represent stable quasiparticles
which undergo correlated tunneling on the lattice \cite{DB9,DB10}. 
Doublon dynamics is experimentally accessible using different platforms, such as ultracold atoms in one-dimensional optical lattices \cite{
DB1,DB9,DB10,DB11,DB12}, 
superconducting quantum metamaterials \cite{DB13} and classical emulators of
two- or three-particle dynamics in Fock space based on two- or three-dimensional photonic \cite{DB14} or topolectrical \cite{DB15,DB16} lattices with engineered defects. 
The current advances in experimental fabrication and control of synthetic matter enables to extend the few-body Hubbard model into the non-Hermitian realm \cite{DB16}, thus motivating the study of 
correlated-particle states and doublon dynamics in non-Hermitian models.\\  
In this work we investigate the spectral and dynamical properties of two strongly-correlated particles on a lattice in an incommensurate sinusoidal potential with a complex phase \cite{r50}, i.e. a non-Hermitian extension of the {\em interacting} Aubry-Andr\'e model \cite{DB11}, 
highlighting distinct spectral and localization-delocalization phase transitions for single-particle and doublon states as the complex phase of the incommensurate potential is varied. Specifically, particle interaction introduces mobility edges which are prevented in the single-particle regime and  lowers the real-to-complex spectral phase transition
 owing to the correlated hopping and slow motion of doublons on the lattice. Remarkably,  since doublons display longer lifetimes than displaced particle states, two particles initially placed at distant sites of the lattice tend to bunch and stick together, forming a doublon state in the long time limit of evolution. This phenomenon, which can be dubbed the {\em non-Hermitian bunching effect}, provides an interesting tool for  "quantum distillation" of interacting particles \cite{distillation} of purely non-Hermitian origin.

\section{Non-Hermitian interacting Aubry-Andr\'e model}
\subsection{Model}
We consider the Hubbard model for interacting fermionic particles in the lowest Bloch band of a one-dimensional tight-binding lattice \cite{BH1} subjected to an external incommensurate sinusoidal potential, which provides an  extension of the Aubry-Andr\'e model for interacting particles [Fig.1(a)]. We indicate by $J$ the single-particle hopping amplitude between adjacent sites in the lattice and by $U$  the on-site interaction energy of fermions with opposite spins ($U>0$ for a repulsive interaction). In the limit $U=0$ and for an Hermitian potential, the model reduces to the ordinary Aubry-Andr\'e model. 
 Non-Hermiticity in the system is introduced by considering a complex phase $\varphi= \theta+ih$ of the sinusoidal on-site potential \cite{r50}, while a reciprocal (Hermitian) amplitude $J$ is assumed for left/right hopping. This means that, contrary to many-body non-Hermitian models considered in several recent works \cite{
I1,I2,I3,I4,I6,I7,I8,I10,I11,I12,I13,I15,I16,I18,I19}, the present model does not display the non-Hermitian skin effect for single-particle states. 
The effective non-Hermitian Hubbard Hamiltonian of the system reads
\begin{eqnarray}
 \hat{H}  & = &   -J \sum_{l, \sigma} \left( \hat{a}^{\dag}_{l, \sigma}  \hat{a}_{l+1, \sigma}+ {\rm H.c.} \right)+ \sum_ {l, \sigma} V_l \hat{n}_{l, \sigma}  \nonumber \\
 &   + &   U \sum_{l} \hat{n}_{l, \uparrow}  \hat{n}_{l, \downarrow}  
\end{eqnarray}
where $\hat{a}_{l, \sigma}$, $\hat{a}^{\dag}_{l, \sigma}$ are the annihilation and creation operators of fermions with spin $\sigma= \uparrow, \downarrow$
 at lattice site $l$, $\hat{n}_{l, \sigma}= \hat{a}^{\dag}_{l, \sigma} \hat{a}_{l, \sigma}$ is the particle-number operator, and
\begin{equation}
V_l=V \cos (2 \pi \alpha l + \varphi)-i \gamma
\end{equation}
is the complex incommensurate on-site potential with real amplitude $V$ and complex phase
\begin{equation}
\varphi=\theta+i h,
\end{equation}
the term $h \geq 0$ governing the strength of non-Hermiticity of the system. The positive constant $\gamma$ provides a uniform loss rate, which just introduces a shift of eigenenergies of the system in the complex energy plane along the imaginary axis and avoids instability in a purely dissipative system. Since the specific value of $\gamma$ does not change the  dynamical behavior of the system under continuous measurements, in the following without loss of generality we will assume $\gamma=0$.
As a typical irrational $\alpha$, we assume the inverse of the golden mean, $\alpha= ( \sqrt{5}-1)/2$, which can be approximated by the sequence of 
rationals $\alpha= \lim_{n \rightarrow \infty} q_{n} / q_{n+1}$, where $q_n$ are the Fibonacci numbers ($q_0=0$, $q_1=1$, $q_{n+1}=q_{n}+q_{n-1}$ for $n \geq 1$). 
In the numerical analysis, we will assume a finite lattice of large size $L=q_{n+1}$ in a ring geometry with periodic boundary conditions ($\hat{a}_{l+L, \sigma}=\hat{a}_{l, \sigma}$) \cite{r50}.
 \begin{figure}[t]
  \centering
    \includegraphics[width=0.4\textwidth]{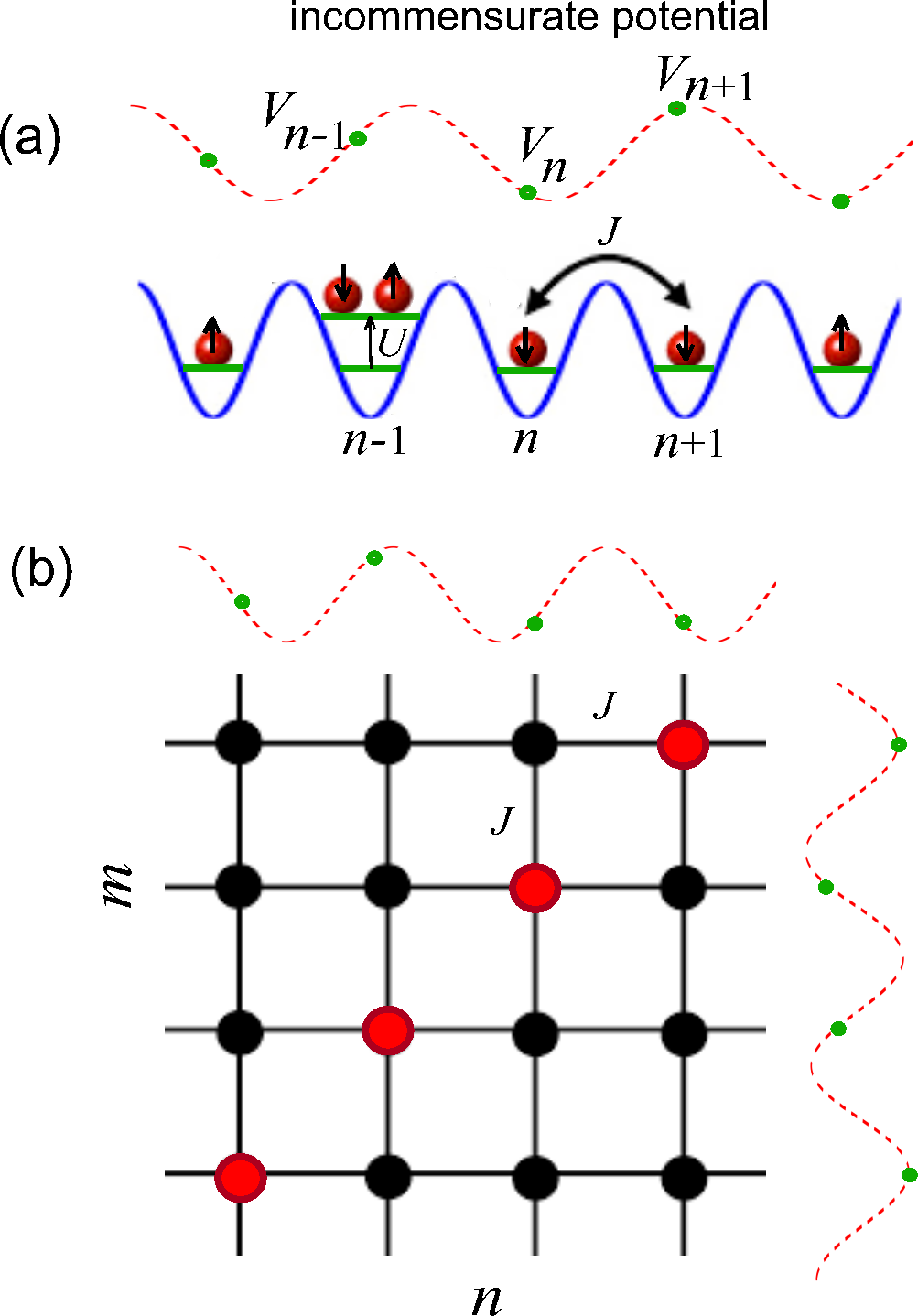}
    \caption{(a) Non-Hermitian Hubbard model, describing correlated many-particle states in a one-dimensional lattice with a superimposed incommensurate on-site potential (quasi crystal). The reciprocal single-particle hopping rate between adjacent sites is $J$, $U$ is the
    on-site interaction energy, and $V_n=V \cos ( 2 \pi \alpha n + \varphi)$ is the incommensurate potential with complex phase $\varphi= \theta+ih$. (b) The Hubbard model for two fermions with opposite spins can be mapped onto the dynamics of a single particle on a square lattice with a line defect on the main diagonal $n=m$ (red circles), corresponding to the interaction energy $U$, and with a hopping rate $J$. States bound on the main diagonal $n=m$ in (b) correspond to sticked two-particle states (doublons) in (a). 
    In the strong interaction regime, doublons undergo correlated hopping via a second-order tunneling process with an effective hopping rate $J_e \simeq 2 J^2/U$.}
     \label{fig1}
 \end{figure}

For pure states and considering open-system dynamics conditioned on
measurement outcomes such that the quantum evolution corresponds to the null-jump process, the state vector $| \psi(t) \rangle$ 
of the system at time $t$ is given by \cite{MB2,MB3,MB4,I2,I18,I19}
\begin{equation}
| \psi(t) \rangle= \frac {\exp(-i \hat{H} t) | \psi(0) \rangle} {\| \exp(-i \hat{H} t) | \psi(0) \rangle \|} \equiv \frac{ | \Phi(t) \rangle}{ \| | \Phi(t) \rangle \|}
\end{equation}
where we have set $| \Phi(t) \rangle \equiv \exp(-iH t) | \psi(0) \rangle$. Basically, at each time interval $dt$ the state vector evolves according to the Schr\^odinger equation with an effective non-Hermitian Hamiltonian $\hat{H}$, followed by a normalization of the wave function, without undergoing any quantum jump. As mentioned above, under post selection the re-normalization of the wave function after each time interval $dt$ makes it the dynamical evolution of $| \psi(t) \rangle$ independent of the loss rate $\gamma$.

\subsection{Single and two-particle states}
The single-particle limit of the Hamiltonian (1), describing the hopping dynamics of a single particle on a one-dimensional incommensurate sinusoidal potential with a complex phase, was earlier introduced Ref.\cite{r50}. In this case, when $V< 2J$ a real-to-complex spectral phase transition, corresponding to a delocalization-localization phase transition, is demonstrated to arise as the non-Hermitian parameter $h$ is increased above the critical value \cite{r50}
\begin{equation}
h_c= \log \left( \frac{2J}{V} \right)
\end{equation}
and can be traced back to the change of a winding number, i.e. to a topological phase transition. For a given base energy $E_B$, which does not belong to the energy spectrum, one can introduce the winding number $w$ \cite{r50}
\begin{equation}
w= \frac{1}{2 \pi i } \int_0^{2 \pi} d \theta  \frac{\partial}{\partial \theta}  \log  \left\{ \det \left( H \left( \frac{\theta}{L}, h \right)-E_B \right) \right\}
\end{equation}
where $H=H( \theta/L,h)$ is the single-particle $L \times L$ matrix Hamiltonian and  $\theta$ is the real phase angle term entering in the incommensurate potential [Eq.(3)]. 
The winding number $w$ counts the number of times the
complex spectral trajectory encircles the base point $E_B$
when the real phase $\theta$ of the potential varies from zero to $ 2 \pi$ \cite{r1}. When the energy spectrum is entirely on the real energy axis, one clearly has $w=0$. Conversely, when the energy spectrum describes one or more closed loops in complex plane, for a base energy $E_B$ internal to one of such loops $w$
takes rather generally a non-vanishing integer value, namely one can show that $w=-1$ independent of $E_B$ \cite{r50}.
Experimental demonstrations of such a kind of non-Hermitian phase transitions in quasi crystals, involving a change of the winding number $w$, have been recently reported in Refs.\cite{r60m,r60n}.\\
In this work we will focus our analysis by considering two fermions with opposite spins hopping on the one-dimensional lattice. In this case, we can expand the state vector $| \psi(t) \rangle$ of the system in Fock space according to
\begin{equation}
| \psi(t) \rangle = \sum_{n,m}  \psi_{n,m}(t) \hat{a}^{\dag}_{n, \uparrow} \hat{a}^{\dag}_{m, \downarrow} | 0 \rangle,
\end{equation}
 Note that the probability of finding the
two particles at the same site $n$ is given by $P_{n}(t ) = |\psi_{n,n}(t )|^2$. 
After letting $\psi_{n,m}(t)= \Phi_{n,m}(t) / \sqrt{ \sum_{n,m} | \Phi_{n,m}(t)|^2 }$,
 the evolution equations for the amplitudes $\Phi_{n,m}(t)$ are readily obtained from the Schr\^odinger equation of a pure state with the effective non-Hermitian Hamiltonian $\hat{H}$ given by Eq.(1), and read
\begin{eqnarray}
i \frac{d\Phi_{n,m}}{dt} & = &    - J \left( \Phi_{n,m+1}+\Phi_{n,m-1}+\Phi_{n+1,m}+\Phi_{n-1,m} \right) \nonumber \\
& + & U \delta_{n,m}\Phi_{n,m}+(V_n+V_m)\Phi_{n,m} 
\end{eqnarray}
Equations (8) show that the hopping dynamics of the two interacting fermions on a one-dimensional lattice is basically equivalent to the hopping motion of a single particle in a two-dimensional square lattice with an incommensurate on-site potential and with an additional defect line on the main diagonal $m= n$, under the periodic boundary conditions $\Phi_{n+L,m}=\Phi_{n,m+L}=\Phi_{n,m}$ [see Fig.1(b)]. Therefore, the spectral and localization properties as well as the dynamical motion of two-particle states in the original one-dimensional quasi crystal can be readily understood by considering the single-particle states in a two-dimensional quasi crystal with a defect line on the main diagonal. Finally, we mention that while our analysis considers two interacting fermions with opposite spins, in the two-particle sector of Hilbert space the Hubbard model Eq.(1) can also describe the dynamical evolution of two identical bosonic particles (rather than two distinct fermions), the only additional constraint being the symmetrization of the wave function under particle exchange. 

\section{Spectral and localization-delocalization phase transitions}
Let us indicate by $\psi_{n,m}^{(\beta, \delta)} \equiv  |\psi^{(\beta, \delta)} \rangle$ and $E_{\beta, \delta}$ the two-particle eigenstates and corresponding eigenenergies of the $L^2 \times L^2$ matrix Hamiltonian corresponding to $\hat{H}$ in the two-particle sector of Hilbert space, where $\beta, \delta=1,2,...,L$ is a pair of indices labeling the matrix eigenstates. 
Here we aim at  exploring the impact of particle interaction on the spectral and localization-delocalization phase transitions found in the single-particle case. For $h>0$, the energy spectrum is rather generally described by energies in complex plane, and a transition from an entirely real to complex spectrum can be detected by monitoring the behavior of $\epsilon \equiv {\rm max}_{\beta, \delta} \left| {\rm Im} \left( E_{\beta, \delta} \right) \right|$ versus the non-Hermitian complex phase $h$.
The localization features of the eigenstates are captured by the inverse participation ratio (IPR), defined by 
\begin{equation}
\text{IPR}^{(\beta, \delta)} =\frac{\sum_{n,m=1}^{L} \left|\psi_{n,m}^{(\beta, \delta)} \right|^{4}} {\left( \sum_{n,m=1}^{L} | \psi_{n,m}^{ ( \beta, \delta )} |^2 \right)^2}
\end{equation}
 The IPR of an extended state scales as $L^{-2}$, hence vanishing in the $L \rightarrow \infty$ limit, while it remains finite for a localized state, with IPR$\leq 1$ and IPR$=1$ when the excitation occupies a single site. In the following, we will indicate by 
IPR$_{max} \equiv {\rm max}_{\beta, \delta}$IPR$^{(\beta, \delta)}$ and 
IPR$_{min} \equiv {\rm min}_{\beta, \delta}$IPR$^{(\beta, \delta)}$ the largest and smallest values of IPR over the eigenstates of the Hamiltonian. For a given base energy $E_B$ that does not belong to the energy spectrum, a winding number $w$ can be introduced, which measures the times the complex two-particle energy spectrum rotates around $E_B$ when the angle $\theta$ adiabatically varies from $0$ to $ 2 \pi$. The definition of $w$ is basically the same as Eq.(6), where now $H( \theta/L, h)$ is the  matrix associated to $\hat{H}$ in the two-particle sector of Hilbert space. Since the energy spectrum in complex energy plane is described by multiple layers of closed loops in complex energy plane [see for example Figs.2(d), 3(d), 4(d) and 5(d) discussed below], whose number increases with the system size $L$, the winding number $w$ is size dependent and can take large values, i.e. not limited to $w=0, -1$ \cite{I7}. For a sufficiently large system size $L$ and $E_B$ not too close to any energy in the spectrum, as previously discussed in Ref.\cite{I7} the function under the sign of the integral in Eq.(6) is almost independent of $\theta$, and thus, after letting 
\[
 {\rm det } \left\{ H \left( \frac{\theta}{L} ,h \right) -E_B \right\}  \equiv R(\theta) \exp [ i \omega (\theta) ],
 \]
 one can simply calculate the winding number using the relation
\begin{equation}
w(E_B) \simeq \frac{d \omega }{d \theta} 
\end{equation}
at any arbitrary value of the angle $\theta$.  It should be mentioned that, while in the single-particle case the winding number is defined taking the $L \rightarrow \infty$ limit \cite{r50}, for two-particle states one should keep the system size $L$ finite since in the limit $L \rightarrow \infty$ the number of closed loops diverges and the energy spectrum covers an entire area (rather than a numerable set of curves).\\
 In the non-interacting limit $U=0$, the two-particle eigenstates $\psi_{n,m}^{(\beta, \delta)}$ and corresponding eigenenergies $E_{\beta, \delta}$ are readily obtained from the single-particle spectral properties of $\hat{H}$, namely one has
\begin{equation}
\psi_{n,m}^{(\beta, \delta)}= \psi_n^{ (\beta)} \psi_m^{( \delta)}
\end{equation}
and
\begin{equation}
E_{\beta, \delta}=E_{\beta}+E_{\delta}
\end{equation}
where $E_{\beta}$ and $\psi_n^{(\beta)}$ are the eigenenergies and corresponding eigenstates of the single-particle Hamiltonian
\begin{equation}
E_{\beta} \psi_n^{(\beta)}=-J \left( \psi_{n+1}^{(\beta)}+\psi_{n-1}^{(\beta)}   \right)+V \cos( 2 \pi \alpha n+\varphi) \psi_n^{(\beta)}.
\end{equation}
\begin{figure}[t]
  \centering
    \includegraphics[width=0.48\textwidth]{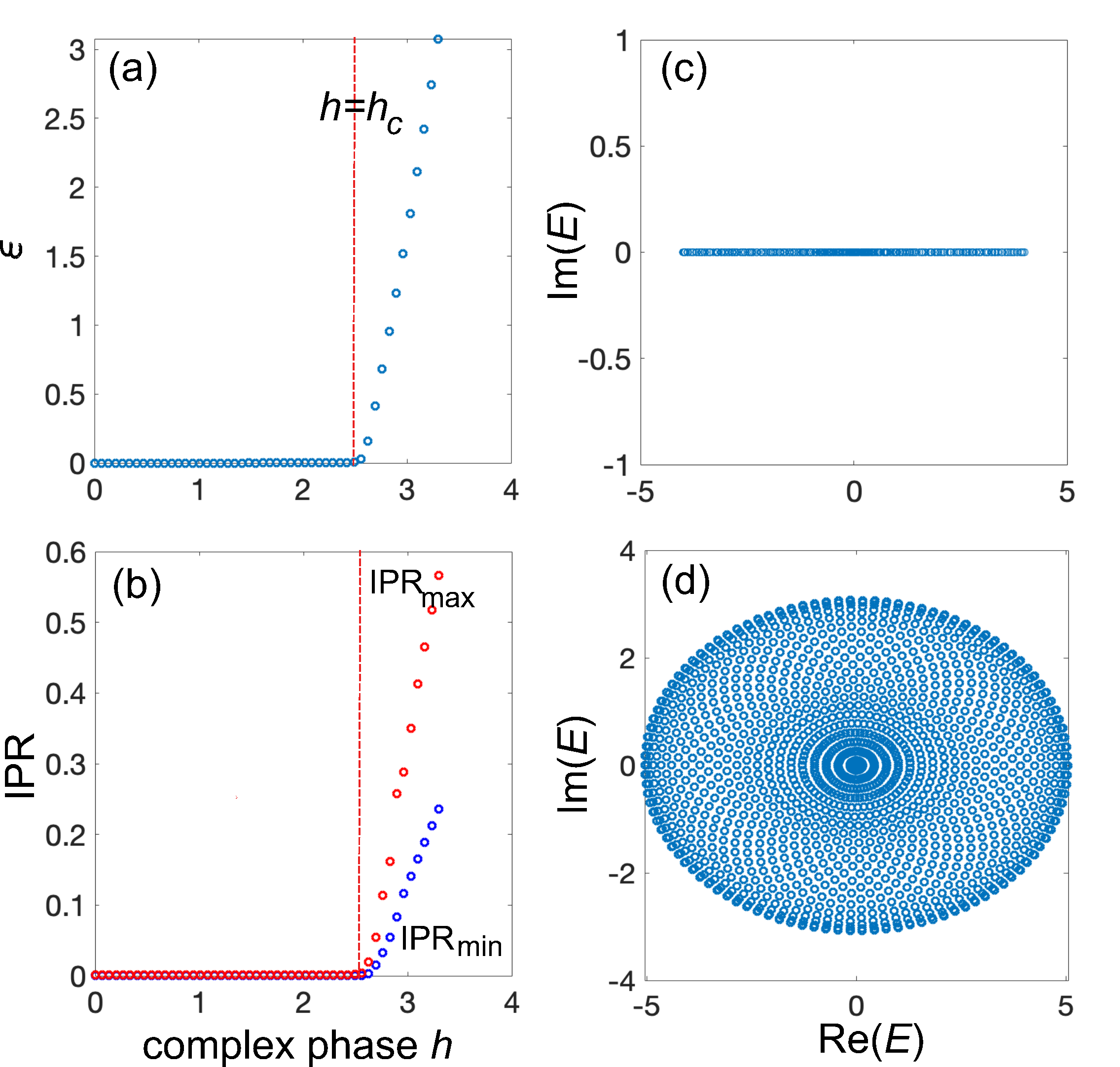}
    \caption{Two-particle energy spectrum and IPR of eigenstates  in the non-interacting limit $U=0$. Other parameter values are $J=1$, $V=0.15$, $\theta=0$, $\alpha=34/55$ and $L=55$. (a) Behavior of the 
    largest value $\epsilon$ (in modulus) of the imaginary part of any eigenenergy versus the complex phase $h$ of the incommensurate potential. The vertical dashed line corresponds to the spectral phase transition point $h=h_c=\log (2J/V) \simeq 2.59$. (b) Behavior of the largest and smallest values of the IPR versus $h$. (c,d) Energy spectra in complex energy plane for $h=1$ [panel (c)] and $h=3.3$ [panel (d)]. In (d) the spectrum comprises multiple layers of closed loops in complex plane, characterized by a non-vanishing winding number $w$ for any base energy $E_B$ internal to such loops. For example,  for the base energies  $E_B=0$, 1.5 and 2.5 one has $w=-55$, -45 and -37,  respectively. }
     \label{fig2}
 \end{figure}
\begin{figure}[t]
  \centering
    \includegraphics[width=0.48\textwidth]{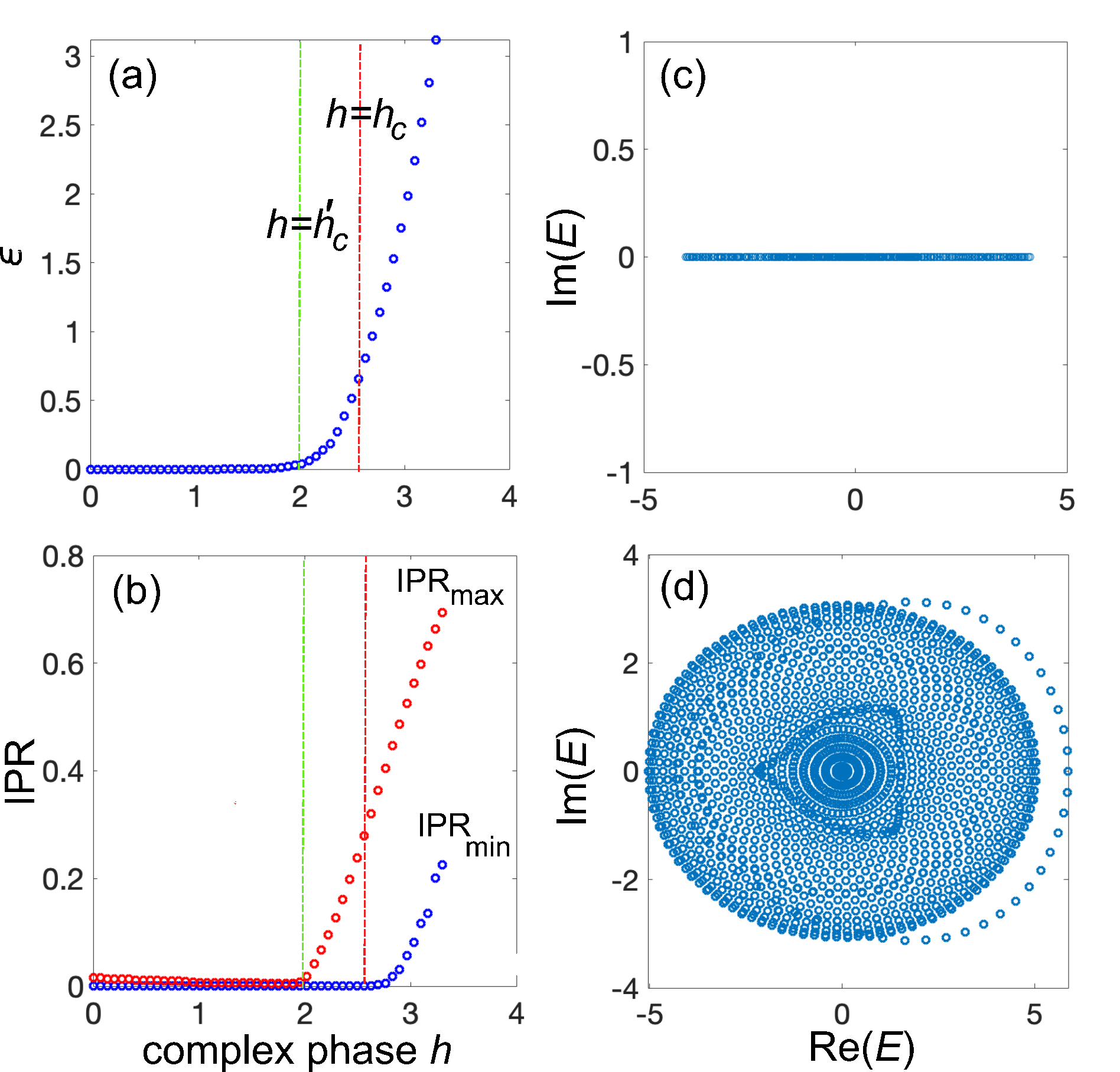}
    \caption{Same as Fig.2, but for $U=1$. Note that the real-to-complex spectral phase transition is observed at a lower value $h_c^{\prime}$ of the complex phase than $h_c$. The value $h_c^{\prime}$ corresponds to the appearance of localized eigenstates, indicated by a finite value of IPR$_{\max}$. For $h<h_c^{\prime}$ all eigenstates are extended, for $h_c^{\prime}<h<h_2$ with $h_2 \simeq h_c$ localized and extended states coexist, whereas for $h>h_2$ all eigenstates are localized.}
     \label{fig3}
 \end{figure}
 \begin{figure}[t]
  \centering
    \includegraphics[width=0.48\textwidth]{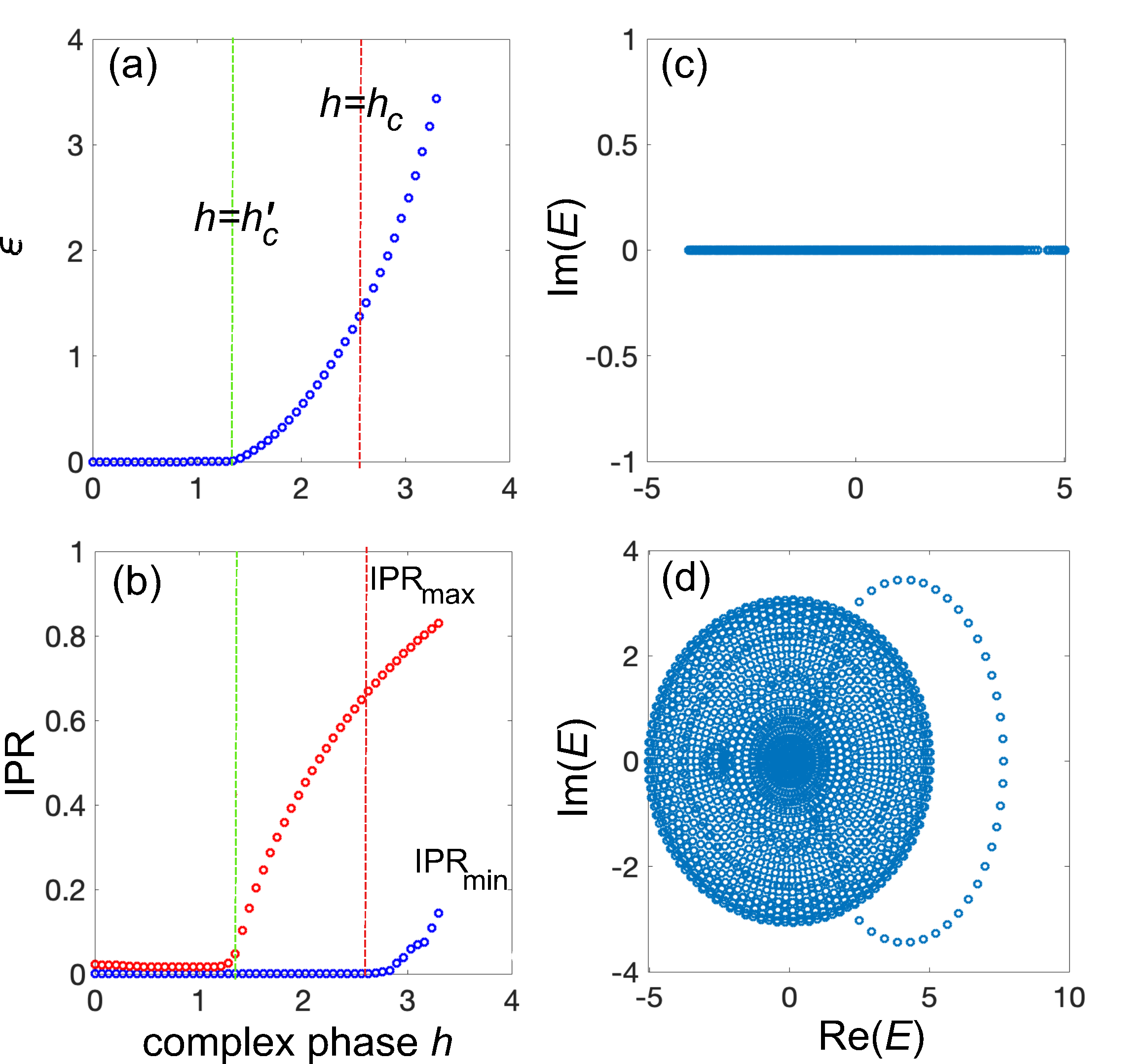}
    \caption{Same as Fig.3, but for $U=3$.}
     \label{fig4}
 \end{figure}
 \begin{figure}[t]
  \centering
    \includegraphics[width=0.48\textwidth]{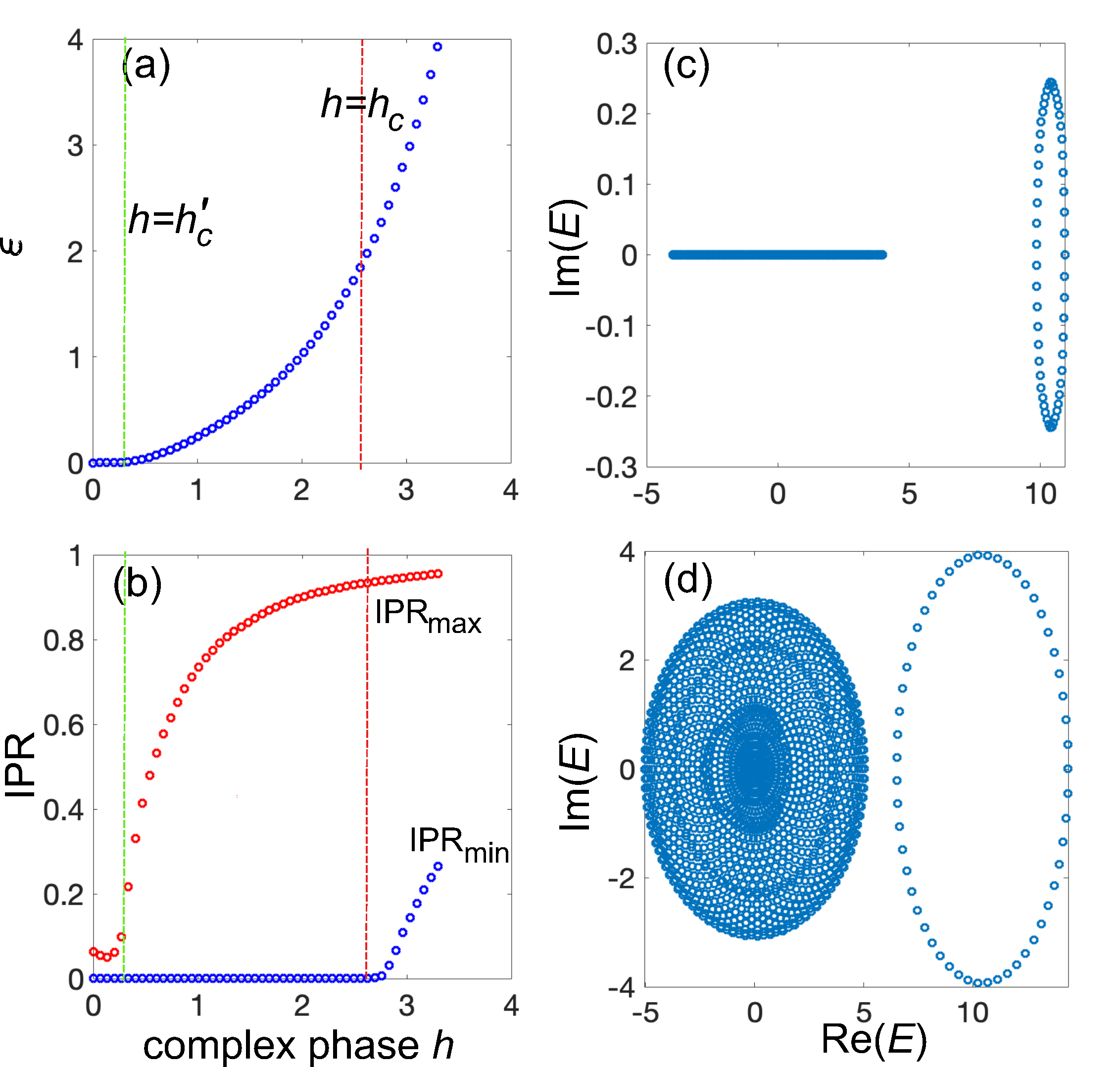}
    \caption{Same as Fig.3, but for $U=10$.}
     \label{fig5}
 \end{figure}
In this case, for $h<h_c$ the energy spectrum remains entirely real and all eigenstates are delocalized, whereas for $h>h_c$ the energy spectrum becomes complex, composed by multiple layers of closed loops in complex energy plane, with simultaneous localization of all corresponding eigenstates.  As an example, Figs.2(a) and (b) show the behavior of $\epsilon$ and of IPR$_{max,min}$ versus $h$, as obtained by numerical diagonalization of the two-particle matrix Hamiltonian, for non-interacting two-particle states in a lattice with parameter values $\alpha=q_{n}/q_{n+1}=34/55$, lattice size $L=q_{n+1}=55$, hopping amplitude $J=1$, potential amplitude $V=0.15$ and phase $\theta=0$. Clearly, for $h<h_c= \log( 2J/V) \simeq 2.59$,  all eigenstates are delocalized and the energy spectrum is real, whereas for $h>h_c$ the energy spectrum becomes complex and all eigenstates become simultaneously localized. Figures 2(c) and (d) show typical energy spectra in the $h<h_c$ and $h>h_c$ phases, respectively. Note that for $h>h_c$ the energy spectrum includes multiple layers of loops in complex plane. Such a layered structure is typical of two-particle states, does not actually require interaction and, as distinctive feature than single-particle states, leads to a winding number $w$
 which is system-size dependent and can take large values, depending on the numbers of loops in the layer (see e.g. \cite{I7}). For example, for the case of Fig.2(d) with $L=55$ the winding number $w$ takes the values $w=-55$, -45 and -37 for a base energy  $E_B=0$, 1.5 and 2.5, respectively.\\
 The spectral and localization properties of the system for increasing values of interaction energy $U$ are shown in Figs.3,4, and 5. The effects arising from particle interaction are mainly twofold. First, one clearly see a lowering 
 of the threshold value $h^{\prime}_c$ for the real-to-complex spectral phase transition than the value $h_c$ predicted in the single-particle case [Eq.(5)].  Second, when the complex phase $h$ varies in the range $h^{\prime}_c< h<h_2$, with $h_2 \sim h_c$, extended and localized states coexist, as one can infer from the inspection of the IPR$_{min}$, which remains close to zero indicating the existence of extended states, and IPR$_{\max}$, which takes finite values corresponding to localized states. Therefore, interaction leads to the appearance of mobility edges, which are prevented in the single-particle (non-interacting) case. Extended and localized states correspond to real and complex energies, respectively. Interestingly, as the interaction energy $U$ increases one loop in the cluster of Fig.2(d) detaches and separates with the formation of a line gap from the other layers, as shown in Figs.3(d), 4(d) and 5(d). For $h>h_2$, deformation of the energy loop layers induced by energy interaction changes rather generally the values of winding numbers $w$ as compared to the non-interacting limit. For example, for the case of Fig.5(d) one has $w=-54,-43,-36$ at the base energies $E_B=0$, 1.5 and 2.5, respectively (to be compared with the values of $w$ given in Fig.1).\\ 
The loop detachment observed as $U$ increases is analogous to the formation of the Mott-Hubbard gap in the standard (Hermitian) Hubbard model in strongly-correlated systems. The detached loop basically describes doublon states, i.e. sticked two-particle states which undergo correlated hopping along the lattice.    
 In the single-particle analogue shown in Fig.1(b), doublon dynamics and correlated particle hopping basically correspond to bound excitations near the defective line along the main diagonal in the square lattice, which cannot spread in the lattice bulk owing to energy conservation constraint. The doublon dynamics can be described in the strong-interaction limit $U \gg J, V\exp(h)$ by a reduced model obtained from a multiple time scale analysis of Eqs.(8), which is detailed in Appendix A. After letting $\Phi_{n,n}(t)=A_n(t) \exp(-iUt)$, one obtains the following evolution equations for the slowly-varying amplitudes $A_n$, corresponding to the amplitude probabilities that the two particles are found at the same lattice site $n$ at time $t$
 \begin{equation}
 i \frac{dA_n}{dt}=J_e (A_{n+1}+A_{n-1}+2A_n)+2V_n A_n
 \end{equation}
 where
\begin{equation}
J_e \equiv \frac{2J^2}{U}
\end{equation}
is effective (second-order) hopping rate of the sticked two-particle state (doublon). Basically, the hopping rate of $J_e$ for doublons, being a second-order process, is greatly reduced as compared to the hopping rate $J$ of a single particle, which corresponds to an effective increase of the amplitude of the incommensurate non-Hermitian potential for doublons. This explains the lowering of the real-to-complex spectral phase transition observed as the interaction energy $U$ is increased, from the single-particle value $h_c$ to the lower value $h^{\prime}_c$, which can be estimated in strong interaction limit using Eq.(14), yielding 
\begin{equation}
h^{\prime}_c \simeq \log \left(
\frac{J_e}{V} \right)= \log \left(
\frac{2J^2}{UV} \right)=h_c-\log \left( \frac{U}{J} \right).
\end{equation}
In particular, when the interaction energy $U$ is  larger than the critical value $U_c=2J^2/V$, one has $h^{\prime}_c=0$, i.e. the energy spectrum becomes immediately complex as soon as the potential is complex. 
\section{Correlated dynamics and non-Hermitian particle bunching} 
In the Hermitian limit of the Hubbard model considered in the previous section two fermions that are initially placed at different sites in the lattice do not tend to bunch and stick together owing to energy repulsion. Likewise, two fermions initially prepared in the same lattice site form a doublon that undergoes correlated hopping on the lattice, i.e. particle dissociation is prevented owing to energy conservation. In the many-particle case where lattice sites can be singly or doubly occupied, the slow motion of doublons as compared to single particle occupancies can be harnessed to realize "quantum distillation", i.e. to separate doublons from singlons \cite{distillation}.\\
The particle dynamics is deeply modified when considering the non-Hermitian extension of the Hubbard model.
 \begin{figure*}[t]
  \centering
   \includegraphics[width=1\textwidth]{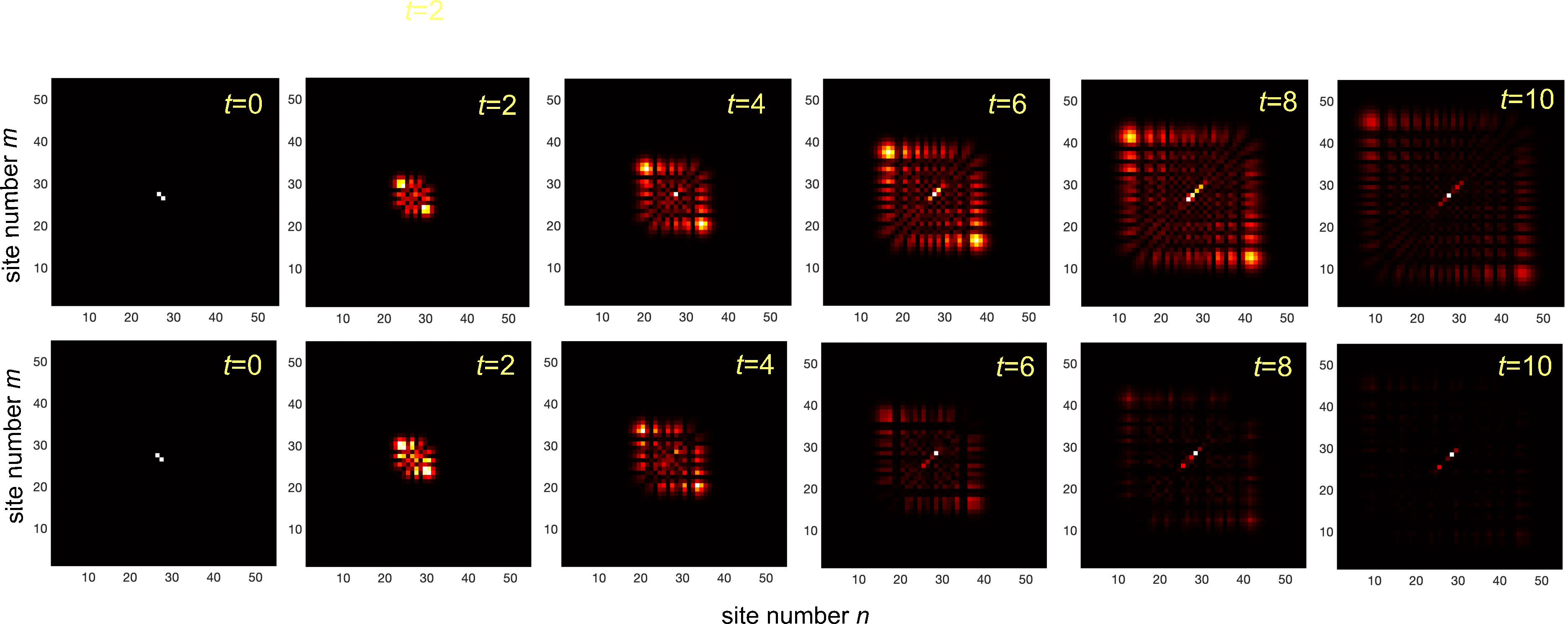}
    \caption{Temporal evolution of the two-particle probability distributions $| \psi_{n,m}(t)|^2$ in a lattice with the same parameter values as in Fig.5 ($J=1$, $U=10$, $V=0.15$, $L=55$, $\theta=0$) and for $h=0$ (Hermitian limit, upper row) and $h=1$ (lower row). The system is initially prepared in a symmetrized state with one particle at site $n_1=26$ and the other one at site $n_2=27$.}
     \label{fig6}
 \end{figure*}
 \begin{figure}[t]
  \centering
    \includegraphics[width=0.5\textwidth]{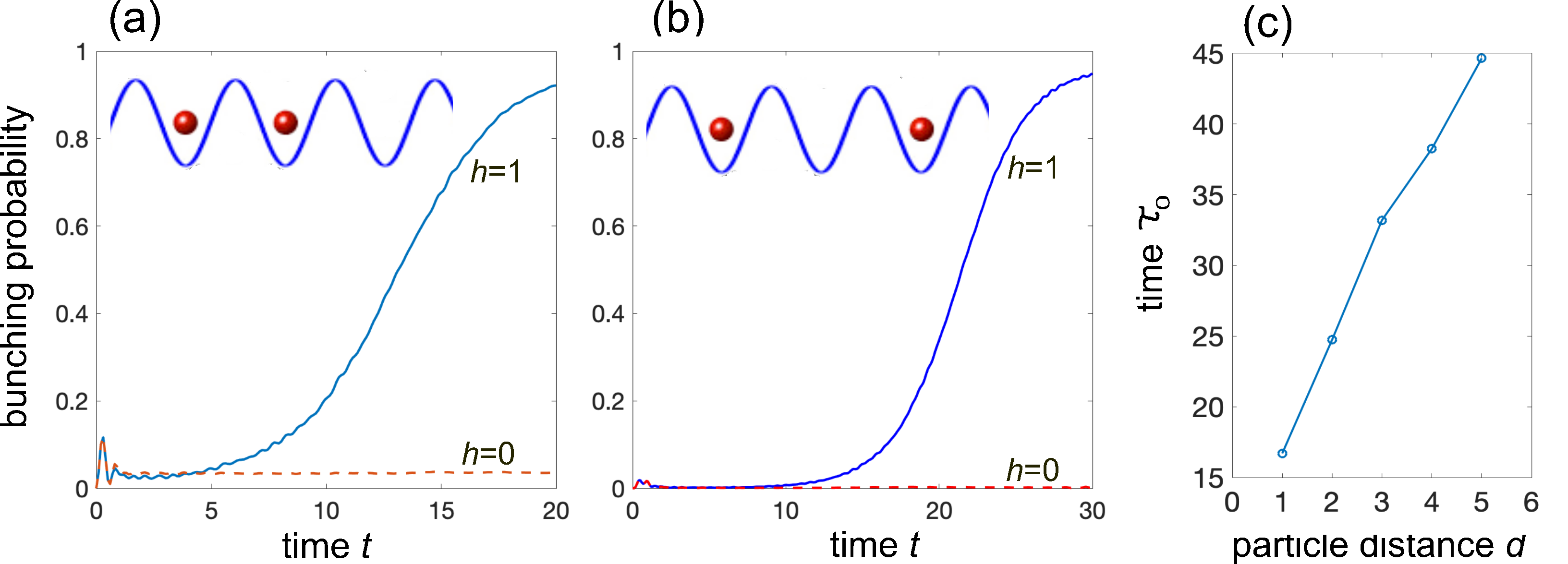}
    \caption{(a) Temporal evolution of the bunching probability  for the same parameter values as in Fig.6 for $h=0$ (Hermitian limit, dotted curve) and $h=1$ (solid curve). (b) Same as (a), bur for an initial separation of the two particles $d=2$. (c) Behavior of the bunching time $\tau_0$ versus initial particle distance $d$ for $h=1$. $\tau_0$ is defined such that $P_{bun}(\tau_0)=0.8$.}
     \label{fig7}
 \end{figure}
In fact, the correlated particle dynamics on the lattice in the non-Hermitian regime $h>h^
{\prime}_c$ is greatly influenced by the different "lifetimes", i.e. imaginary parts of  eigenenergies, of doublon states  than single-particle states. Since doublons display longer lifetimes, in the strong interaction regime a rather arbitrary initial excitation of the system, that is not  exactly orthogonal to doublon eigenstates, is attracted toward one of such eigenstates in the long time limit, i.e. the two particles tend to bunch and stick together as a result of the non-Hermitian dynamics, a phenomenon that can be referred to as {\em non-Hermitian particle bunching}. We note such a phenomenon is distinct than  quantum distillation of doublons and singlons observed in the Hermitian case and briefly mentioned above,  where single-particle states move faster than doublons and can escape from the edges of the system \cite{distillation}.\\ 
To understand the non-Hermitian bunching effect, let us observe that for a given initial condition the state of the system at time $t$ can be written as
\begin{equation}
| \psi(t) \rangle = \frac
{ \sum_{\beta, \delta} C_{\beta,\delta} | \psi^{ (\beta,\delta)}  \rangle \exp(-iE_{\beta, \delta} t)}
{  \| \sum_{\beta, \delta} C_{\beta,\delta} | \psi^{ (\beta,\delta)}  \rangle \exp(-iE_{\beta, \delta} t)  \|}
\end{equation}
 where the spectral amplitudes $C_{\beta,\delta}$ are determined by the initial state $ | \psi(0) \rangle$ and are given by
 \begin{equation}
C_{\beta, \delta}= \langle \psi^{ \dag (\beta, \delta)} | \psi(0) \rangle. 
\end{equation}
In the above equation,  $\psi^{ \dag (\beta, \delta)}$  are the eigenfunctions of the adjoint Hamiltonian $\hat{H}^{\dag}$ in the two-particle sector of Hilbert space, which is obtained from $\hat{H}$ by just reversing the sign of $h$, and the orthonormal conditions $ \langle \psi^{ \dag (\beta, \delta)} | \psi^{ (\beta', \delta' ) } \rangle= \delta_{\beta, \beta'} \delta_{\delta, \delta'}$ are assumed. From Eq.(17), it then follows that the long-time dynamics of the system is dominated by the excited two-particle eigenstate of $\hat{H}$ with the largest imaginary part of the eigenenergy, which is expected to be rather generally a doublon eigenstate.\\ 
As an example, let us assume that at initial time $t=0$ the two particles are initially placed at sites $n_1$ and $n_2$ of the lattice, distant one another by $d=|n_2-n_1|$, and let us consider the symmetrized wave function as an initial state
\begin{equation}
| \psi(0) \rangle= \frac{1}{\sqrt{2}} \left( \hat{a}^{\dag}_{n_1,\uparrow} \hat{a}^{\dag}_{n_2,\downarrow} | 0 \rangle +\hat{a}^{\dag}_{n_2,\uparrow} \hat{a}^{\dag}_{n_1,\downarrow} | 0 \rangle  \right)
\end{equation}
(symmetrization of the wave function is assumed so as to include in the analysis the case of two bosonic particles as well).  The probability that at time $t$ the two particles stick together (bunching probability) is computed from the relation
\begin{equation}
P_{bun}(t) = \sum_{n} | \psi_{n,n}(t)|^2.
\end{equation}
Figure 6 shows the numerically-computed temporal evolution of the site occupation probabilities of the two particles on the lattice, initially placed at a distance $d=1$ one another, for the same parameter values as in Fig.5 ($J=1$, $L=55$, $\alpha=34/55$, $V=0.15$, $\theta=0$) in the Hermitian ($h=0$, upper row) and non-Hermitian ($h=1$, lower row) regimes, clearly indicating that in the latter case the two particles tend to stick together, while in the former case they do not. The corresponding behavior of the bunching probability is shown in Fig.7(a). 
The time required for the two particles to bunch together is clearly dependent on the lifetime differences between doublon and single-particle eigenstates and on the weight $|C_{\beta, \delta}|$ of the overlapping of initial state onto the doublon eigenstate with the largest growth rate, which decreases as the particle distance $d$ increases. Hence, the time of the two particles to stick together increases as $d$ is increased; compare e.g. Figs.7(a)  and 7(b), where the initial distance $d$ of the two particles is increased from $d=1$ to $d=2$. Quantitatively, we can define a bunching time $\tau_0$ such that $P_{bun}(\tau_0)$ reaches a target  (reference) value, for example 80\%. A typical behavior of $\tau_0$ versus initial particle separation $d$ is shown in Fig.7(c), indicating that $\tau_0$ increases almost  linearly with $d$.




\section{Conclusions}
Topological phases and phase transitions in non-Hermitian crystalline or quasi-crystalline systems provide a fascinating area of research with promising implications in different fields of physics, from condensed matter to cold atoms and classical systems such as photonic, acoustic, mechanical and topolectrical settings. While the properties of single-particle non-Hermitian models have been the subject of extensive studies and revealed unprecedented phenomena without any counterpart in Hermitian systems, such as the appearance of non-trivial point-gap and line-gap topologies, the non-Hermitian skin effect, an extended form of the bulk-boundary correspondence and a variety of dynamical and transport effects, 
intriguing physical phenomena are being  discovered when considering interacting many-body non-Hermitian systems. 
In this work we investigated the spectral and dynamical features of two interacting particles in a non-Hermitian quasi crystal, described by an effective Hubbard model in an incommensurate sinusoidal potential with a complex phase,  and 
unravelled some intriguing effects without any Hermitian counterpart.  Owing to an effective increase of disorder strength introduced by particle interaction, doublon states, i.e. bound particle states, display a much lower threshold for spectral and localization-delocalization transitions than single-particle states, leading to the formation of interaction-induced mobility edges. Remarkably, since doublons display longer lifetimes, two particles initially placed at distant sites in the lattice tend to bunch and stick together, forming a doublon state in the long time limit of evolution, a phenomenon that can be dubbed {\em non-Hermitian particle bunching}. 
Our results shed new light onto the physical properties of strongly correlated particles in non-Hermitian systems, even in the few-body case considered here, and could 
suggest novel possibilities to control many-particle states harnessing non-Hermitian physics. In the present study, the analysis has been focused on a specific non-Hermitian interacting Aubry-Andr\'e model, where non-Hermiticity enters via a complex phase $h$ in the incommensurate on-site potential, however different non-Hermitian versions of the interacting Aubry-Andr\'e model could be considered, such as the Aubry-Andr\'e model with off-diagonal incommensurate disorder \cite{Referee1,Referee2} or the Aubry-Andr\'e model with non-reciprocal (asymmetric) hopping amplitudes \cite{Referee3,Referee4,Referee5} induced by an imaginary magnetic flux $\eta$ \cite{Referee6},  which displays the non-Hermitian skin effect in the single-particle regime \cite{
r5,r5c,r5h}. In particular, it would be interesting to investigate the interplay and competition between complex phase  $h$ of the incommensurate potential, which tends to localize the wave functions, and the magnetic flux $\eta$, which tends to delocalize the wave functions and plays the same role as $h$ but in reciprocal (Fourier) space \cite{r50}. Since the imaginary magnetic flux $\eta$ acts in a different way for two-particle scattered states and two-particle bound states (doublons) \cite{Referee7}, the non-Hermitian bunching effect is expected to be washed out by the imaginary magnetic flux and doublon dissociation in the bulk would be observed for a magnetic flux $\eta$ large enough than the complex phase $h$.


\acknowledgments
The author acknowledges the Spanish State Research Agency, through the Severo Ochoa
and Maria de Maeztu Program for Centers and Units of Excellence in R\&D (Grant No. MDM-2017-0711).\\
\\
This work is dedicated to my beloved mother,  Anna, who passed away recently.

\appendix
\section{Doublon dynamics: multiple-time scale analysis}
In this Appendix we derive Eqs.(14) and (15) given in the main text, which describe the correlated hopping of two particles in the lattice that occupy the same site at each time (doublons). 
Let us introduce the normalized time variable $\tau=Ut$, so that Eqs.(8) take the form
\begin{eqnarray}
i \frac{d\Phi_{n,m}}{d \tau} & = &    - \frac{J}{U} \left( \Phi_{n,m+1}+\Phi_{n,m-1}+\Phi_{n+1,m}+\Phi_{n-1,m} \right) \nonumber \\
& + &  \delta_{n,m}\Phi_{n,m} +\frac{V_n+V_m}{U} \Phi_{n,m}.
\end{eqnarray}
Let us now consider the strong interaction regime by assuming $J/U \equiv \epsilon$, with $\epsilon \ll 1$. We also consider a potential strength $V$ and complex phase $h$ such that $U \gg V \exp(h)$, with 
$V \exp(h) /U \sim \epsilon^2$. Therefore, in Eq.(A1) the first term on the right hand side of the equation is of order $\sim \epsilon$, the second term is of order $\sim \epsilon^0$, and the last term is of order $\sim \epsilon^2$. Let us now assume that at initial time the system is prepared in a doublonic state, i.e. such that $\Phi_{n,m}( \tau=0)=0$ for $ n \neq m$, and let us look for a solution to Eq.(A1) as an asymptotic series
\begin{equation}
\Phi_{n,m} (\tau) = \Phi_{n,m}^{(0)} (\tau)+ \epsilon \Phi_{n,m}^{(1)} (\tau)+ \epsilon^2 \Phi_{n,m}^{(2)}(\tau)+...
\end{equation}
To ensure that the expansion (A2) is uniformly valid as $\tau$ grows, multiple time scales
\begin{equation}
T_0= \tau \; , \;\;  T_1= \epsilon \tau \; , \;\; T_2= \epsilon^2 \tau \; ,\;\;  ...
\end{equation}
have to be introduced to avoid the occurrence of secular growing terms in the asymptotic expansion.  Using the derivative rule
\begin{equation}
\frac{d}{d \tau}= \frac{d}{dT_0}+ \epsilon \frac{d}{dT_1}+ \epsilon^2 \frac{d}{d T_2}+...
\end{equation}
substitution of Eqs. (A2) and (A4) into Eq. (A1), and after
equating terms of the same power in $\epsilon$, a hierarchy of equations
for successive corrections to $\Phi_{n,m}$ is obtained. At leading order $ \sim \epsilon^0$  one simply obtains
\begin{equation}
i \frac{\partial \Phi_{n,m}^{(0)}}{\partial T_0}= \delta_{n,m} \Phi_{n,m}^{(0)}
\end{equation}
which can be readily solved by letting
\begin{equation}
\Phi_{n,m}^{(0)}= A_n \delta_{n,m} \exp(-i T_0)
\end{equation}
where the amplitudes $A_n$ can vary on the slow time scales $T_1$, $T_2$,..., i.e. $A_n=A_n(T_1,T_2,...)$. At order $\sim \epsilon$ one has
\begin{equation}
i \frac{\partial \Phi_{n,m}^{(1)}}{\partial T_0}= -\left( \Phi_{n+1,m}^{(0)}+  \Phi_{n-1,m}^{(0)} + \Phi_{n,m+1}^{(0)} + \Phi_{n,m-1}^{(0)}\right)
\end{equation}
for $n \neq m$, and 
\begin{equation}
i \frac{\partial \Phi_{n,n}^{(1)}}{\partial T_0}- \Phi_{n,n}^{(1)}= -i \frac{\partial A_n}{\partial T_1} \exp(-i T_0).
\end{equation}
The solvability condition for Eq.(A8) yields
\begin{equation}
\frac{\partial A_n}{\partial T_1} =0 \; ,\;\;\; \Phi_{n,n}^{(1)}=0,
\end{equation}
whereas Eq.(A7) can be solved for $\Phi_{n,m}^{(1)}$ ($n \neq m$), yielding
\begin{eqnarray}
\Phi_{n,m}^{(1)} & = & - \left( A_{n+1} \delta_{m,n+1} +A_{n-1} \delta_{m,n-1}  \right. \\
& + & \left.  A_{n} \delta_{m,n+1}+A_{n} \delta_{m,n-1} \right)  \exp(-i T_0) \nonumber
\end{eqnarray}
Finally, at order $\sim \epsilon^2$ for $n=m$ one obtains
\begin{eqnarray}
i \frac{\partial \Phi_{n,n}^{(2)}}{\partial T_0}- \Phi_{n,n}^{(2)} & = &  \exp(-i T_0) \left\{  -i \frac{\partial A_n}{\partial T_2} + \frac{2 V_n}{ \epsilon^2 U} A_n\right.  \nonumber \\
& + & \left. 2 \left( A_{n+1}+A_{n-1}+2 A_n \right)    \right\}.
\end{eqnarray}
The solvability condition to Eq.(A11) yields
\begin{equation}
i \frac{\partial A_n}{\partial T_2} = \frac{2 V_n}{ \epsilon^2 U} A_n-+2 \left( A_{n+1}+A_{n-1}+2 A_n \right).
\end{equation}
 If we stop the asymptotic analysis at order $\sim \epsilon^2$, after reintroduction of the original variables from Eqs. (A2), (A4), (A6), (A9) and (A12) one finally
obtains
 \begin{eqnarray}
\Phi_{n,n}(t) & = & A_n(t)\exp(-iUt) +O(\epsilon^2)  \nonumber \\
\Phi_{n,m}(t) & = & O(\epsilon) \;\; (m = n \pm 1) \\
\Phi_{n,m}(t) & = & o(\epsilon) \;\; ( |m-n| \geq 2) \nonumber
\end{eqnarray}
where the slowly-varying amplitudes $A_n$ evolve according to the following equations
\begin{equation}
i \frac{dA_n}{dt}=\frac{2J^2}{U}(A_{n+1}+A_{n-1}+2A_n)+2 V_n A_n
\end{equation}
which correspond to Eqs.(14) and (15) given in the main text, with $J_e \equiv 2 J^2/U$. Note that $J_e$  corresponds to an effective hopping rate of two-particle states, which is a second-order process (as it arises at order $\sim \epsilon^2$ in the asymptotic analysis).

\end{document}